\documentclass[12pt]{scrartcl}
\usepackage{amsfonts,epsf,graphicx,mathbbol,verbatim}
\usepackage{epsfig}

\def\ha{\frac{1}{2}}
\def\=!{\stackrel{(!)}{=}}

\def\pa{\partial}
\def\D{\mbox{D}}
\def\d{\mbox{d}}

\def\eps{\epsilon}

\def\>{\rightarrow}

\begin{document}

\title{\begin{flushright}
\normalsize FSU TPI 07/00
\end{flushright}
\vspace{2cm}
\bf \huge Magnetic monopoles vs. Hopf defects
 in the Laplacian (Abelian) gauge}
\vspace{2cm}

\author{F.~Bruckmann, T.~Heinzl\thanks{~~Supported by DFG.},~
        T.~Vekua\thanks{~on leave from 
        Department of Physics, Tbilisi State University,
        Charchavadze Avenue 3, 380028 Tbilisi, Georgia;
        address after July 15th: Institut f\"ur
        Theoretische Physik, Universit\"at Hannover, Appelstra\ss{}e
        2, D-30167 Hannover.}~~and
        A.~Wipf \\
        Friedrich-Schiller-Universit\"at Jena\\ 
        Theoretisch-Physikalisches Institut \\ 
        Max-Wien-Platz 1, D-07743 Jena\\}

\date{\today}


\maketitle

\thispagestyle{empty}

\begin{abstract}
We investigate the Laplacian Abelian gauge on the sphere $S^4$ in the
background of a single `t~Hooft instanton. To this end we solve the
eigenvalue problem of the covariant Laplace operator in the adjoint
representation.  The ground state wave function serves as an auxiliary
Higgs field. We find that the ground state is always degenerate and has
nodes. Upon diagonalisation, these zeros induce toplogical defects in
the gauge potentials. The nature of the defects crucially depends on
the order of the zeros. For first-order zeros one obtains magnetic
monopoles. The generic defects, however, arise from zeros of second
order and are pointlike. Their topological invariant is the Hopf index
$S^3\rightarrow S^2$. These findings are corroborated by an analysis
of the Laplacian gauge in the fundamental representation where similar
defects occur. Possible implications for the confinement scenario are
discussed.

\end{abstract}

\newpage

\section{Introduction}

Although not derived from first principles,
the dual superconductor scenario
\cite{thooft:76a,mandelstam:76,parisi:75} is
widely believed to explain color confinement in QCD.  To realise this
idea, `t~Hooft suggested to use Abelian projections \cite{thooft:81a}
which allow for a straightforward identification of magnetic monopoles
in pure Yang Mills theories.  In this approach one fixes the gauge
group up to its maximal Abelian subgroup. This partial gauge fixing
can be characterised by a Higgs field $\phi$ in the adjoint
representation, which becomes diagonal in the Abelian gauge
(AG)\footnote{In the following we will distinguish between the
Abelian gauge, which is a partial gauge fixing, and the Abelian
projections, where one neglects the off-diagonal part of
the gauge field after gauge fixing.}.
Magnetic monopoles arise as gauge fixing defects
whenever $\phi$ vanishes. At these points, the gauge transformation
diagonalising $\phi$ becomes ambiguous. In the low temperature phase
of QCD these defects should condense and play the role of Cooper
pairs.

This picture is strongly supported by lattice calculations (for recent
reviews, see \cite{chernodub:98,haymaker:99}). In the continuum,
however, Abelian gauges are not that well understood. Considerable
progress has only been made for the Pol\-ya\-kov Abelian gauge (PAG)
\cite{weiss:81,reinhardt:97b,jahn:98,ford:98,ford:99a,ford:99b}.
The  defects occuring in the PAG are characterised by a winding number
$S^2\rightarrow S^2$ of the (normalised) Higgs field, $n \equiv
\phi/|\phi|$,  or equivalently by the magnetic charge $q$ of the
Abelian gauge field.  A relation between monopole charge $q$ and
instanton number $\nu[A]$ has been established which enforces the
presence of monopoles in any non-trivial instanton sector ($\nu \ne
0$).

For the maximally Abelian gauge (MAG)
\cite{thooft:81a,kronfeld:87a,kronfeld:87b}, there are only few
analytical results. It is known that configurations with monopole
lines \cite{chernodub:95} and monopole loops
\cite{brower:97b} are in this gauge. They are, however, strongly
suppressed by the gauge fixing functional, at least in the backgound
of single instantons.  Recently, it has been explicitly shown that the
continuum MAG suffers from a Gribov problem \cite{gribov:78} as
expected from Singer's theorem \cite{singer:78}: the `t~Hooft
instanton in the singular gauge is located on the Gribov horizon of
the MAG \cite{bruckmann:00}.

In order to circumvent the Gribov (`spin glass') problem of the MAG on
the lattice, the Laplacian Abelian gauge (LAG) has been proposed as a
superior alternative
\cite{vandersijs:96,vandersijs:98a,vandersijs:98b}.  Some first
applications of this idea in the context of lattice gauge theory have
appeared only recently \cite{alexandrou:99a,alexandrou:99b}.
Analytically, however, it seems that only one result has been obtained
so far: by comparing the behaviour of the gauge fixing functionals one
finds \cite{vandersijs:96} that in the LAG magnetic degrees of freedom
are less suppressed than in the MAG. Some deeper understanding of this
Abelian gauge is obviously desirable.

This paper presents our first investigation of the continuum Laplacian
Abelian gauge. In order to have a large amount of symmetry, we
consider (the orbit of) a single 't~Hooft instanton.  As we shall
see, the LAG is somewhat ill-defined on infinite-volume manifolds, and
thus we compactify space to a sphere $S^4$. For a special choice of
the compactification radius, the symmetry is enhanced to SO(5) so that
the (gauge fixing) problem can be exactly solved. For other radii,
symmetry arguments still provide some insights. Going back to infinite
volume, we find that the singular gauge instanton and global SU(2)
rotations thereof lie in the LAG. Accordingly, the instanton in the
singular gauge is a horizon configuration, as was the case for the
MAG.

Of particular interest is the question how the submanifolds of
vanishing Higgs field look like. It has been argued that, generically,
these are loops, i.e.~closed monopole worldlines, having
codimension three\footnote{as one has to solve three equations on a
four-dimensional manifold}. The instanton number can be recovered
from these loops for general Higgs fields
\cite{jahn:00,tsurumaru:00}. It is related to the winding number
$S^2\rightarrow S^2$. While the same is true for the LAG, we find that
the Higgs field associated with a single 't~Hooft instanton in
addition induces \textit{pointlike} defects, i.e.~events localised in
space-time. The corresponding
topological invariant is the Hopf index $S^3\rightarrow S^2$.

This paper is organised as follows: First, in Section~2, we define the
LAG and discuss its properties. Single 't~Hooft instantons on $S^4$
are introduced in Section~3.  In Section~4 we diagonalize the
covariant Laplacian in the adjoint representation. A classification of
its ground state wave functions, which serve as auxiliary Higgs
fields, is given in Section~5. A brief discussion of the Laplacian
gauge (corresponding to the fundamental representation) is added in
Section~6. Finally, we conclude with some remarks on the physical
implications of our findings.

\section{The Laplacian Abelian gauge}

The Laplacian Abelian gauge on $\mathbb{R}^4$ is defined by minimising
the Higgs kinetic energy \cite{vandersijs:98a,vandersijs:98b}, 
\begin{eqnarray}
\label{funct}
  F_{\rm LAG}[A,\phi]=\ha\int
  (\D_\mu\,\phi^a\D_\mu\,\phi^a-E\phi^a\phi^a)\:\d^4x,\qquad
  \D_\mu=\pa_\mu-i[A_\mu,\ldots] \; , 
\end{eqnarray}
with respect to the auxiliary Higgs field $\phi$ in the adjoint
representation.  The energy variable $E$ is a Lagrange multiplier
demanding that $\phi$ is square integrable, $\int \phi^a\phi^a \:
\d^4x < \infty$. The field configuration $\phi$ minimising
$F_{\mathrm{LAG}}$ can be viewed as the ground state of the covariant
Laplacian,
\begin{eqnarray}
\label{eigen}
  - \D_\mu^2[A] \phi = E \, \phi,  
\end{eqnarray}
where $E$ is the ground state energy. Obviously, (\ref{eigen})
represents a four dimensional Schr\"odinger problem with a potential
essentially given by $A^2$.

The gauge transformation $\Omega$ diagonalising $\phi$ puts the gauge
field $A$ into the LAG,
\begin{eqnarray}
  A_{\rm LAG}\equiv\,^\Omega\!A,\qquad
  \mbox{where }\:\:^\Omega\!\phi\equiv\Omega^{-1}\phi\,\Omega\sim\sigma_3.
\end{eqnarray}
$\Omega$ may be ambiguous for two reasons. First, if $\phi$ has zeros
and second, if the groundstate is degenerate. One can use node and
uniqueness theorems  to analyse these issues \cite{reed:78}.


On a space-time with infinite volume, the LAG is not straightforwardly
defined for the following reasons.  Since $-D_\mu^2[A]$ is a
non-negative operator we have $E \geq 0$.  Moreover, whenever the
gauge field $A$ tends to zero at infinity, there are scattering states
and the continuous spectrum always starts from zero.  Scattering
states, however, are not normalisable.  Thus, for a generic background
(including the instantons to be studied), one does not expect that the
covariant Laplacian $-D_\mu^2[A]$ will have a normalisable ground
state. The situation is quite analogous to the quantum mechanics of
the ordinary `Laplacian' $d^2 / dx^2$ on the real line.  We avoid this
problem by considering gauge fields on the four-sphere $S^4$ which
leads to a purely discrete spectrum of the associated covariant
Laplacian.

\section{Single instanton on the sphere}

In the following we  consider the single 't~Hooft
instanton (in singular and regular gauge) on a sphere $S^4$ of
radius $R$.  On Euclidean $\mathbb{R}^4$ the configurations read,
\begin{eqnarray}
\label{inst}
  A_\mu^{\rm sg}=\bar{\eta}^a_{\mu\nu}x_\nu
  \frac{\rho^2}{r^2(r^2+\rho^2)}\sigma_a,\qquad
  A_\mu^{\rm reg}=\eta^a_{\mu\nu}x_\nu
  \frac{1}{(r^2+\rho^2)}\sigma_a,\qquad
  r^2 = x_\mu x_\mu \; ,
\end{eqnarray}
using the conventions of \cite{schaefer:98}. The configurations
(\ref{inst}) are related by the gauge transformation
\begin{eqnarray}
\label{h}
  h=\hat{x}_4\Eins_2+i\hat{x}_a\sigma_a,\qquad
  \hat{x}_\mu\equiv x_\mu/r \; , 
\end{eqnarray}
which also relates the solutions $\phi$ of (\ref{eigen}) in these
two backgrounds, $\phi^{\rm reg}=h\,\phi^{\rm sg}\,h^\dagger$. 

We benefit from the facts that classical Yang-Mills theories are
conformally invariant and that the sphere $S^4$ is conformally
equivalent to compactified Euclidean space $\dot{\mathbb{R}}^4$.  If
we use conformal coordinates $x_\mu$ on the sphere, which are simply
the Cartesian coordinates of the point stereographically projected
onto $\mathbb{R}^4$, the metric is conformally flat,
\begin{eqnarray}
\label{metric}
  g_{\mu\nu}(x) = e^{\alpha_R(r)}\delta_{\mu\nu} \equiv
  \frac{4R^4}{(r^2+R^2)^2} \, \delta_{\mu\nu} \; .
\end{eqnarray} 
Field configurations minimising the Yang Mills action on
$\mathbb{R}^4$ are also minimising configurations on the sphere, if
the Cartesian coordinates are substituted by conformal
coordinates. Thus, we can simply use expressions (\ref{inst}) for the
instantons on the sphere.

What about the symmetry of these configurations? It is known
\cite{Jackiw:80b} that on $\mathbb{R}^4$ they are invariant under
SO(4) rotations and a combination of translations and special
conformal transformations,
\begin{equation}
  \delta x_\mu = \omega_{\mu\nu}x^\nu+2\,c\cdot
  x\,x_\mu/\rho-c_\mu(x^2+\rho^2)/\rho \; , 
\end{equation}
up to a compensating gauge transformation, $\delta A_\mu =
\D_\mu\theta$, with 
\begin{equation} 
  \theta^a=\ha\omega^{\mu\nu}\eta^a_{\mu\nu}-2c^\mu\eta^a_{\mu\nu}x^\nu\qquad
  \mbox{(reg. gauge).}
\end{equation}
Together these transformations form (a non-linear representation of)
the group SO(5). This symmetry is preserved on $S^4$ when the radius
$R$ of $S^4$ coincides with the instanton size $\rho$
\cite{Jackiw:76b}. To illustrate this point, we note that the gauge
invariant Lagrangian density,
\begin{eqnarray}
  \mathcal{L}\sim
  g^{\mu\rho}g^{\nu\sigma}\:
  F_{\mu\nu}^a F_{\rho\sigma}^a
  =\frac{12\rho^4}{R^8}\:\frac{(r^2+R^2)^4}{(r^2+\rho^2)^4},
\end{eqnarray}
is constant on $S^4$ (and thus SO(5)-invariant) only if $R=\rho$.  For
$R\neq\rho$ the explicit appearance of $r$, which is only
SO(4)-invariant, breaks SO(5) down to SO(4).

Since compensating gauge transformations do not spoil equation
(\ref{eigen}), the eigenfunctions $\phi$ furnish representations of
SO(5) and SO(4), respectively.

\section{Solutions of the covariant Laplacian}

Generalising equations (\ref{funct}) and (\ref{eigen}) to
curvilinear coordinates, we define the LAG on the sphere $S^4$ via the
functional
\begin{eqnarray}
  F_{\rm LAG}[A,\phi]=\ha\int_{S^4} \left(\D_\mu\phi^a\D_\nu\phi^ag^{\mu\nu}-
  E\phi^a\phi^a\right)\sqrt{g}\,\d^4x \; , 
\end{eqnarray}
where $g = \exp(2\alpha_R)$ denotes the determinant of the metric
(\ref{metric}).  The equation of motion is given in terms of the
(gauge covariant) Laplace-Beltrami operator,
\begin{eqnarray}
  -\frac{1}{\sqrt{g}}\D_\mu\sqrt{g}g^{\mu\nu}\D_\nu\phi=E\phi. 
\end{eqnarray}
To proceed we make use of the symmetry and separate into angular and
radial equations.  The  angular part is expressed in terms of angular
momenta derived from the decomposition  $so(4)\cong su(2)\oplus
su(2)$,
\begin{eqnarray}
  M_a=-\frac{i}{2}\bar{\eta}^a_{\mu\nu} x_\mu \pa_\nu,\:\:
  \vec{M}^2\rightarrow m(m+1),\quad
  N_a=-\frac{i}{2}\eta^a_{\mu\nu} x_\mu \pa_\nu,\:\:
  \vec{N}^2\rightarrow n(n+1) \; .
\end{eqnarray}
In this representation, the two Casimirs coincide, $\vec{M}^2 =
\vec{N}^2$. Their eigenvalues are half-integer,
$m=n\in\{0,\,1/2,\,1,\,3/2,\ldots\}$. The generators for isospin $t =
1$ are 
\begin{eqnarray}
  (T_a)_{bc}=i\eps_{bac},\qquad \vec{T}^2\rightarrow t(t+1) = 2 \; .
\end{eqnarray}
The radial equations on the sphere differ from those
in Euclidean space by a metric factor, $\exp (- \alpha_R)$, and  a
dilatation term, $r\pa_r$, 
\begin{eqnarray}
  e^{-\alpha_R(r)}
  \left[-\pa_r^2-\frac{3}{r}\pa_r+\frac{4 \vec{M}^2}{r^2}+
  \frac{4\rho^2(\vec{J}^2-\vec{M}^2)}{r^2(r^2+\rho^2)}-
  \frac{4\vec{T}^2\rho^2}{(r^2+\rho^2)^2}+
  \frac{4r}{r^2+R^2}\pa_r
  \right]\phi^{\rm sg}
  =E\phi^{\rm sg} \label{eigeneqn1} \\
  e^{-\alpha_R(r)}
  \left[-\pa_r^2-\frac{3}{r}\pa_r+\frac{4 \vec{N}^2}{r^2}+
  \frac{4(\vec{J}^2-\vec{N}^2)}{(r^2+\rho^2)}-
  \frac{4\vec{T}^2\rho^2}{(r^2+\rho^2)^2}+
  \frac{4r}{r^2+R^2}\pa_r
  \right]\phi^{\rm reg}
  =E\phi^{\rm reg} \label{eigeneqn2}
\end{eqnarray}
In the above, we have introduced the conserved angular momentum
$\vec{J}$ (`spin from isospin',
\cite{jackiw:76c,hasenfratz:76,goldhaber:76}),
\begin{eqnarray}
\label{jdef}
  \vec{J}\equiv\vec{L}+\vec{T},
  \qquad\vec{J}^{\,2}\rightarrow j(j+1),
  \qquad j\in\{l-1,l,l+1\} \; , 
\end{eqnarray}
where $\vec{L}$ denotes $\vec{M}$ or $\vec{N}$, respectively.
Replacing angular momenta by their eigenvalues and exchanging
$j\rightarrow n,\,m\rightarrow j$, equation (\ref{eigeneqn1}) turns
into (\ref{eigeneqn2}). This amounts exactly to the action of the
gauge transformations $h$ from (\ref{h}).  

The symmetry considerations above suggest the following form of the
ground state,
\begin{eqnarray}
  \phi(x)=Y_{(j,l)}(\hat{x})\varphi(r) \; ,
\end{eqnarray}
where the $Y$'s denote the spherical harmonics on $S^3$ (see
Appendix~\ref{spherharm}). Note that there are two competing angular
momentum terms in (\ref{eigeneqn1}) and (\ref{eigeneqn2}), so that it
is not obvious in which angular momentum sector the groundstate will
be.  By simply looking at the radial potentials in the different
sectors, we can only state the following bound on the energy in an
arbitrary sector,
\begin{eqnarray}
\label{groundstates}
  E_{(j,l)} \geq \min \{E_{(0,1)},\,E_{(1/2,1/2)},\,E_{(1,0)}\} \; .
\end{eqnarray}
The quantum numbers of the ground state candidates on the r.h.s.\
correspond to the representations (0,1), (1/2, 1/2) and (1,0) of
$su(2)_j\oplus su(2)_l$ and thus have degeneracies 3, 4 and 3,
respectively.  Note that the singlet $(0,0)$ is excluded by the
selection rules for $t=1$, see (\ref{jdef}). Accordingly, for any of
the possible choices in (\ref{groundstates}), the groundstate will be
degenerate. The spherical harmonics for the three cases are listed in
Appendix~\ref{spherharm}.

At this point two further remarks are in order: First, the radial part
$\varphi$ shows power law behaviour in $r$, both for small and large
$r$, independent of $R$ and $\rho$,
\begin{eqnarray}
  \varphi^{\rm sg}(r\rightarrow0)\rightarrow r^{2j} \; , &&\qquad
  \varphi^{\rm sg}(r\rightarrow\infty)\rightarrow r^{-2m} \; ,
  \label{asympt1} \\
  \varphi^{\rm reg}(r\rightarrow0)\rightarrow r^{2n} \; , &&\qquad
  \varphi^{\rm reg}(r\rightarrow\infty)\rightarrow r^{-2j} \; . 
  \label{asympt2} 
\end{eqnarray}
Second, upon substituting $\varphi \equiv (r^2+R^2)\cdot\chi$ and
$\lambda \equiv ER^2+2$, one can absorb the dilatation term,
\begin{eqnarray}
  \left[-\pa_r^2-\frac{3}{r}\pa_r+\frac{4 \vec{M}^2}{r^2}+
  \frac{4\rho^2(\vec{J}^2-\vec{M}^2)}{r^2(r^2+\rho^2)}-
  \frac{4\vec{T}^2\rho^2}{(r^2+\rho^2)^2}-
  \frac{4\lambda R^2}{(r^2+R^2)^2}
  \right]\chi^{\rm sg}=0 \; ,
  \label{eqn}\\
  \left[-\pa_r^2-\frac{3}{r}\pa_r+\frac{4 \vec{N}^2}{r^2}+
  \frac{4(\vec{J}^2-\vec{N}^2)}{(r^2+\rho^2)}-
  \frac{4\vec{T}^2\rho^2}{(r^2+\rho^2)^2}-
  \frac{4\lambda R^2}{(r^2+R^2)^2}
  \right]\chi^{\rm reg}=0 \; . 
\end{eqnarray}
Setting $R=\rho$, the differential equation (\ref{eqn}) coincides with
the one considered by 't~Hooft in his analysis of the fluctuations
around instantons \cite{thooft:76c}.  The eigenvalues are
$\lambda_k=(k+j+l+1-t)(k+j+l+t+2)$.  The lowest energy corresponds to
$k=0$ and $j+l=1$, consistent with the three possible groundstates of
(\ref{groundstates}). Together they form the 10-dimensional adjoint
representation\footnote{Using the conventions of
\cite{cornwell:84}, this representation is labelled by the
integers $\{n_1,n_2\}=\{0,2\}$ which are the coefficients of
the highest weight when expanded in terms of the fundamental weights.}
of SO(5) \cite{cornwell:84}.  The value of the ground state energy is
$E=2/R^2$.

\begin{figure}[t!]
\begin{center}
\epsfig{figure=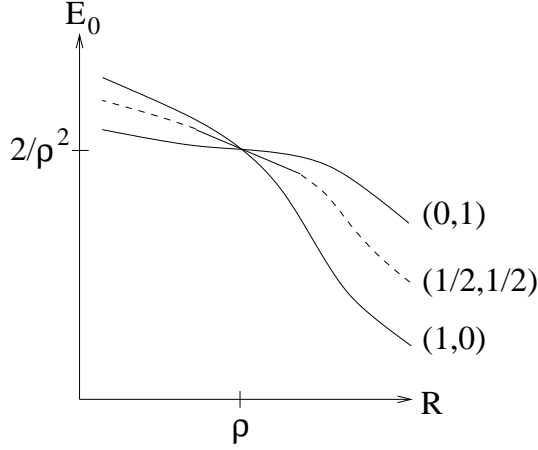,scale=0.7}
\end{center}
\caption{Energy of the lowest-lying states in the relevant
angular momentum sectors as functions of the compactification radius
$R$ (singular gauge). At the point $R=\rho$ the two triplets and 
the quadruplet meet, while for $R\rightarrow\infty$ the triplet
$(1,0)$ has lowest energy. For symmetry reasons we expect the
dashed line to stay inbetween the other two for $R\neq\rho$.}
\label{er}
\end{figure}

The radial eigenfunctions are rational,
\begin{eqnarray}
  \label{radial}
  \varphi^{\rm sg}(r) = R \, \frac{(r/R)^{2j}}{r^2+R^2} \; , \qquad
  \varphi^{\rm reg}(r) = R \, \frac{(r/R)^{2-2j}}{r^2+R^2} \; , 
\end{eqnarray}
and obey the asymptotics (\ref{asympt1}) and (\ref{asympt2}),
respectively. In accordance with the node theorem for the
one-dimensional radial equation, the lowest-lying states have no zeros
apart from $r=0$ and $r = \infty$.

For the cases $R>\rho$ and $R<\rho$ we cannot solve the radial
equation analytically.  However, we are able to prove the following
statements:

For the singular gauge and $R > \rho$, the triplet $(1,0)$ has lower
energy than the triplet $(0,1)$. For $R < \rho$, the situation is vice
versa with the triplet $(0,1)$ having lower energy. Analogous
statements hold for the regular gauge (see Fig.~\ref{er}). These
results are a straightforward consequence of the Feynman-Hellmann
theorem (cf. Appendix~\ref{FH}).

For the quadruplet (1/2, 1/2), the situation is somewhat more
complicated. Using perturbation theory in $\delta=\rho^2-R^2$ (see
Appendix~\ref{FH}), one finds that these states have energy inbetween
the two disjoint triplet states. For symmetry reasons we do not expect
the spectral flow $E_{(1/2,1/2)} (R)$ to intersect the others for some
$R
\ne\rho$ (see Fig.~\ref{er}). 

Finally, the node theorem again guarantees that $\phi$ vanishes only
at $r=0$ and $r = \infty$, in accordance with the asymptotics
(\ref{asympt1},\ref{asympt2}).

\section{Properties of the solutions}

\begin{figure}
\begin{center}
\epsfig{figure=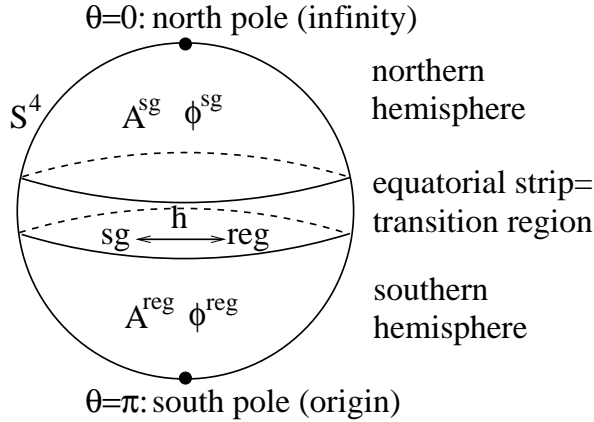,scale=0.7}
\end{center}
\caption{In the bundle picture on $S^4$ there are two gauge and Higgs
fields, which are smoothly defined on their domains (hemispheres). In
the transition region, they are related by the gauge transformation
(transition function) $h$. Note that the four-dimensional radius can
be expressed in terms of the azimuthal angle $\theta$,
$r=R\cot(\theta/2)$.}
\label{bundle}
\end{figure}

Before characterising the zeros of the solutions $\phi$, let us point
out the following subtlety: Near the origin, the (0,1) wave functions
(Higgs fields) in the singular gauge are bilinear in $\hat{x}_\mu$ and
thus \textit{discontinuous} there. They inherit this singularity from
the instanton field\footnote{which results in the asymptotics
$\varphi(r) \sim r^0$, see (\ref{asympt1}).}. Nevertheless, the wave
functions are square integrable on $S^4$ due to the measure factor
$r^3$. The same, of course, is true for the regular gauge states near
infinity. In order to work with smooth Higgs fields, it is appropriate
to use the \textit{principal fibre bundle} picture. This can be viewed
as a non-Abelian $S^4$-analogue of the Wu-Yang construction for the
Dirac monopole on $S^2$ \cite{wu:75}.  The $A$-field in the regular
gauge represents the connection smoothly\footnote{ to be precise: The
$A$-fields are the pullbacks of the connection under smooth sections
of the bundle.} on the southern hemisphere (the chart containing the
origin), while the $A$-field in the singular gauge does the same on
the northern hemisphere (the chart containing infinity). In the
transition region formed by the equatorial strip displayed in
Fig.~\ref{bundle}, the gauge transformation $h$ from (\ref{h})
interpolates between the two.  For simplicity we will retract the
transition region to a single three-sphere $S^3_r$ of fixed
four-dimensional radius $r$ (fixed azimuthal angle $\theta$).

The Higgs field is a section in an associated fibre bundle: on each of
the two charts there is a Higgs field. In the transition region, the
same transition function $h$ relates the two (see Fig.~\ref{bundle}).
Our results obtained so far can immediately be
carried over to the bundle picture, since, for every solution in the
singular gauge, there is a corresponding gauge transformed `mirror'
solution in the regular gauge with the same energy (and vice versa).
Moreover, the angular momenta are interchanged by $h$ in such a way that the
\textit{radial} wavefunctions (\ref{radial}) are smooothly defined on
the whole of $S^4$.  The complete eigenfunctions $\phi$
are continuous in their
respective charts but `jump' (in their isospin direction) due to the
action of the transition function $h$ in the transition region.\\

Along these lines, let us discuss the ground state in the $(1/2,1/2)$
sector, which has zeros localised on \textit{loops}.  To simplify the
discussion, we choose  the fourth of the spherical harmonics in
(\ref{harmonics12_sg}) or (\ref{harmonics12_rg}), for which
\begin{eqnarray}
  \phi^{\rm sg}(x)
  =\left(\begin{array}{c}x_1\\x_2\\x_3\end{array}\right)
  \frac{1}{r^2+R^2}
  =\phi^{\rm reg}(x),
\end{eqnarray}
since $h$ from (\ref{h}) commutes with $\phi^{\rm sg}$.
This Higgs field $\phi$ is of hedgehog type. Thus, its diagonalisation
induces a Dirac monopole at $\vec{x}=0$ and a Dirac string along
the negative 3-axis. The world-line of the monopole is the great 
circle in 4-direction (which degenerates to the 4-axis in the
infinite-volume limit). For the other three states in the multiplet
the same holds true upon permutation of the coordinates.

It is well-known \cite{arafune:75} that the monopole charge is
characterised by the winding number of the \textit{normalised} Higgs
field $n=\phi/|\phi|$, explicitly given by
\begin{eqnarray}
  n^{\rm sg}(x)
  =\left(\begin{array}{c}x_1\\x_2\\x_3\end{array}\right)
  /|\vec{x}|
  =n^{\rm reg}(x) \; .
\end{eqnarray}
The $n$-field is singular at $\vec{x}=0$, where $\phi$ vanishes. This
singularity has the following topological characterisation. Consider
the two-sphere, $S^2$: $\vec{x} = const.$, surrounding the
singularity. There, the $n$-field provides a smooth mapping,
$S^2\rightarrow S^2 \cong SU(2)/U(1)$, labelled by an integer, the
winding number, which in our case is just one.

\begin{figure}
\begin{minipage}{0.45\linewidth}
\begin{center}
\epsfig{figure=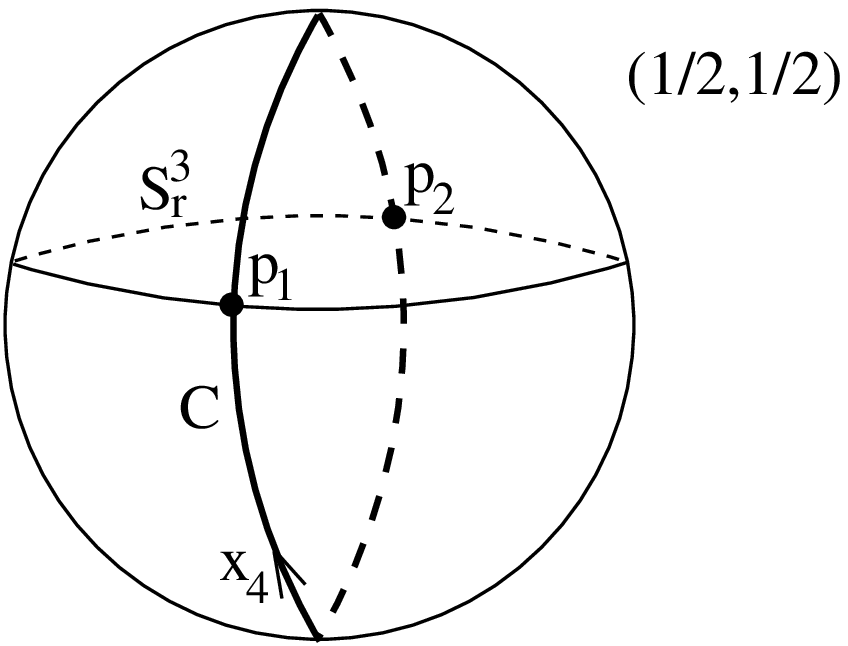,scale=0.7}
\end{center}
\end{minipage}
\begin{minipage}{0.45\linewidth}
\begin{center}
\epsfig{figure=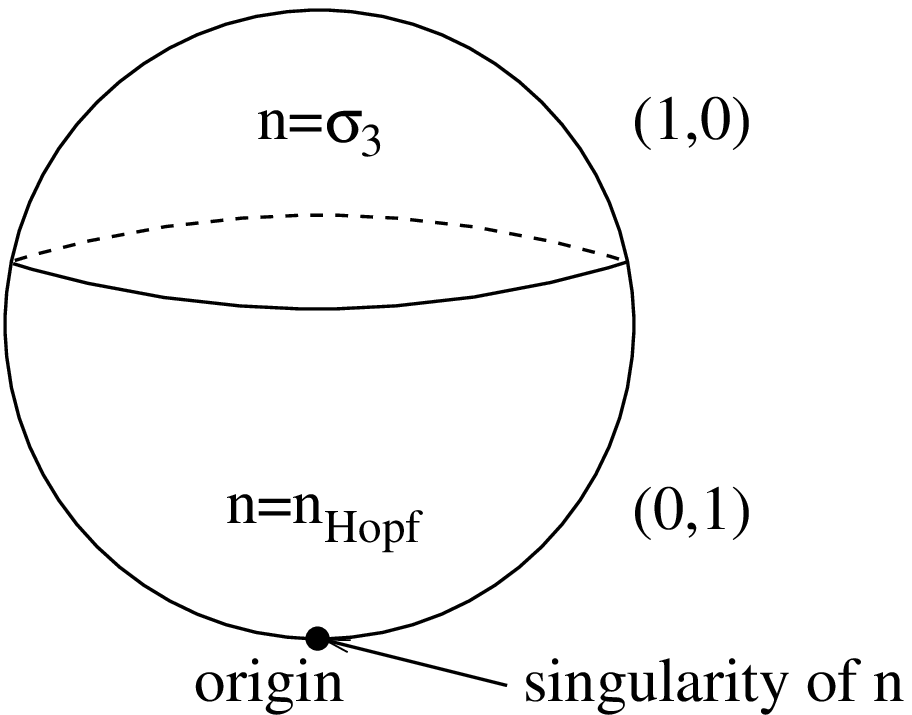,scale=0.7}
\end{center}
\end{minipage}
\caption{Submanifolds of $S^4$ on which the ground state wave
function vanishes so that the normalised Higgs field becomes
singular. The sector (1/2,1/2) gives rise to monopole loops $C$, while
the generic sectors (0,1) and (1,0) lead to pointlike singularities
with Hopf index as topological invariant.}
\label{loops}
\end{figure}

The LAG Higgs field also serves as an illustration of the relation
between instanton number and monopole charge recently proposed in
\cite{tsurumaru:00}.  We note that in the (1/2, 1/2) sector
the two Higgs fields $n^{\rm sg}$ and $n^{\rm reg}$ coincide on the
whole of $S^4$, their singularities being located at two points
$p_1=(\vec{0},-r),\,p_2=(\vec{0},r)$ in the retracted transition
region $S^3_r$ (see Fig.~\ref{loops}).  Consider the submanifold
$S^3_r \, \setminus \, \{p_1,\,p_2\} \cong S^2\times I_{12}$. In terms
of the polar angle $\vartheta\in(0,\pi)$ on the three-sphere, the
intervall $I_{12}$ is parametrised by $x_4=r\cos\vartheta\in(-r,r)$,
while the two-spheres are given by $|\vec{x}|=r\sin\vartheta$.

We already know that the magnetic charge measured on the two-sphere is
$q=1$. If we express the transition function $h$ in terms of
$n$ and $\vartheta$, 
\begin{eqnarray}
  h=\exp(i\vartheta  \, n^a\sigma_a) \; , 
\end{eqnarray}
the flux\footnote{notice the difference $\sigma_a$
vs. $\sigma_a/2$ as compared to \cite{tsurumaru:00}}
$\Phi = \int_{I_{12}} \d(2\vartheta)$ is  easily computed as
\begin{eqnarray}
  \Phi=2\int_0^\pi\d\vartheta=2\pi \; .
\end{eqnarray}
We thus recover the instanton number,
\begin{eqnarray}
  \nu[A]=q\,\frac{\Phi}{2\pi} = 1 \; . 
\end{eqnarray}
Note that for linear combinations of states from different multiplets
the monopole loops become \textit{tilted}. As an example, take a
combination of the sectors (1,0) and (1/2, 1/2),
\begin{eqnarray}
  \phi^{\rm reg}=\frac{1}{\sqrt{2}}\left(\phi^{\rm reg}_{(1,0)}-
  \phi^{\rm reg}_{(1/2,1/2)}\right)
  =\frac{1}{\sqrt{2}(r^2+R^2)}
  \left(\begin{array}{c} 
  x_1\\
  x_2\\
  R-x_3
  \end{array}\right) \; .
\end{eqnarray}
This Higgs field vanishes for $x_\mu=(0,0,R,x_4)$, a set of zeros
which is still a great circle but does no longer include the poles.
 
As we have already argued, the quadruplet states (1/2, 1/2) will occur
as ground states only for $R=\rho$. In the general case, $R \ne \rho$,
the ground states will be the triplet states $(0,1)$ and $(1,0)$,
which have \textit{isolated, pointlike} zeros. Let us specialise to
the physical region $R>\rho$ which includes the infinite-volume limit,
$R\rightarrow\infty$.  (For $R<\rho$ one has to perform the
appropriate `mirror' transformation.)  We know that for the southern
hemisphere (regular gauge) and for the northern hemisphere (singular
gauge) we have to take $(j,l)=(0,1)$ and $(j,l)=(1,0)$, respectively,
since these multiplets consist of the lowest-lying states. If we
choose the third of the spherical harmonics in (\ref{harmonics10}) and
(\ref{harmonics01_rg}), the normalised Higgs field is given by
\begin{eqnarray}
  n(x)=\left\{\begin{array}{ll}
  n^{\rm reg}(x)=
  \left(\begin{array}{c}
  2(\hat{x}_1\hat{x}_3-\hat{x}_2\hat{x_4})\\
  2(\hat{x}_2\hat{x}_3+\hat{x}_1\hat{x_4})\\
  -\hat{x}_1^2-\hat{x}_2^2+\hat{x}_3^2+\hat{x}_4^2
  \end{array}\right)
  &\mbox{ southern hem.}\ni 0 \; ,\\
  n^{\rm sg}(x)=
  \left(\begin{array}{c}0\\0\\1\end{array}\right)
  &\mbox{ northern hem.}\ni \infty \; .
  \end{array}\right.
\end{eqnarray}
$n^{\rm reg}$ is singular at the origin $r=0$ and closely resembles
the standard Hopf map \cite{Hopf:31,Nakahara:90}.  For any finite
radius $r\neq 0$, it provides a smooth mapping $S^3_{r={\rm
fixed}}\rightarrow S^2\cong SU(2)/U(1)$ with Hopf index one
\cite{ryder:80}.

As a result, we have obtained the simplest realisation of the
connection between instanton number and Hopf indices  derived in
\cite{jahn:00}: The (signed) sum of all Hopf indices of $n$ around its
singularities equals the instanton number $\nu$. This statement is
analogous to results from residue calculus where the singularities of
$n$ (the zeros of $\phi$) are replaced by the poles of a meromorphic
function: The (signed) sum of all residues equals the residue at
infinity.
Like the magnetic monopoles in the PAG, $n$ must possess
singularities in any non-trivial instanton sector ($\nu \ne 0$). In
addition, pairs of singularities may occur which do not contribute to
the instanton number.

Coming back to the LAG, the remaining task is to diagonalise the
ground-state Higgs field $\phi(x)$. On the northern hemisphere, where
$\phi$ is already diagonal, there is nothing to be done.  No gauge
transformation is needed and the gauge field remains in the singular
gauge. On the southern hemisphere, we basically have to diagonalise
the standard Hopf map. This is achieved by the gauge transformation
$h$, which transforms the gauge field $A$ from regular to singular
gauge. Independent of where we choose the transition region,
\textit{the LAG-fixed configuration on the orbit of the single
't~Hooft instanton is in the singular gauge} (for $R>\rho$). Notice
that the gauge fixed configuration inherits a singularity only at the
point where $n$ is singular; there are no further `Dirac strings'.

If we choose an arbitrary linear combination of the triplet spherical
harmonics, the diagonalising gauge transformation includes an
additional \textit{global} SU(2) rotation. Together with $A^{\rm sg}$,
all its global rotations are located on the gauge fixing hypersurface
defined by the LAG. We thus find a whole $S^3$ of gauge-equivalent
configurations (Gribov copies).

\section{The Laplacian gauge}

The importance of pointlike defects (as compared to loops) is
corroborated by their occurence in a closely related gauge, the
Laplacian gauge (LG)\footnote{The authors thank P.~de Forcrand for
drawing their attention to this issue.}. The Laplacian gauge
\cite{schierholz:85,vink:92,vink:95,vanbaal:95} is defined via a Higgs
field $q$ in the \textit{fundamental} representation, being the ground
state of the covariant Laplacian,
\begin{eqnarray}
  -\D_\mu^2[A]q=E\,q \; , \qquad
  \D_\mu=\pa_\mu-iA_\mu \; .
\end{eqnarray} 
It is a \textit{complete} gauge fixing (up to defects) if the
two-component complex vector $q$ is rotated into a fixed isospin
direction and made real,
\begin{eqnarray}
  ^\Omega\!q\equiv\Omega^{-1} q={|q| \choose 0} \; , \qquad  
  A_{\rm LG}\equiv\,^\Omega\!A \; .
\end{eqnarray}
Our formalism is easily adapted to this gauge by choosing the isospin
$t=1/2$ representation in terms of the Pauli matrices
$T_a=\sigma_a/2$. For $R=\rho$ one again has to minimise $j+l$,
whence $(j,l)=(0,1/2)$ or $(j,l)=(1/2,0)$. As before, these states
form an irreducible representation\footnote{the four-dimensional spinor
representation labelled by $\{n_1,n_2\}=\{0,1\}$ in
\cite{cornwell:84}.} of SO(5).  For $R>\rho$ and the singular gauge, 
the state $(1/2,0)$ has lowest energy (by the same Feynman-Hellmann
argument) so that the singular gauge instanton again satisfies the
gauge condition.  The relevant spherical harmonics are
\begin{eqnarray}
  Y^{\rm reg}_{(1/2,0)}&=&\left\{
  \left(\begin{array}{c}1\\0\end{array}\right),\:
  \left(\begin{array}{c}0\\1\end{array}\right)
  \right\}\; ,\\ 
  Y^{\rm reg}_{(0,1/2)}&=&\left\{
  h\left(\begin{array}{c}1\\0\end{array}\right)=
  \left(\begin{array}{c}\hat{x}_4+i\hat{x}_3\\
  -\hat{x}_2+i\hat{x}_1\end{array}\right),\:
  h\left(\begin{array}{c}0\\1\end{array}\right)=
  \left(\begin{array}{c}\hat{x}_2+i\hat{x}_1\\
  \hat{x}_4-i\hat{x}_3\end{array}\right)
  \right\} \; , 
\end{eqnarray}
which are nonzero throughout $S^4$. In analogy with (\ref{asympt2}),
we have the following behaviour near the origin, $q(r)\sim r^{2n}=r$
for ($j$, $n$) = (0, 1/2). Thus, the modulus of the Higgs field is
proportional to the four-dimensional distance $r$ from the origin
(where the topological charge of the instanton is concentrated). This
perfectly agrees with latest results from lattice simulations
\cite{deforcrand:00}.

Again, a topological description is possible. On a three-sphere
surrounding the origin, one can define $n \equiv
q/|q|:\,S^3\rightarrow S^3$ with integer winding number. In the case
above, the $n$-field simply reduces to the identity map, 
\begin{equation}
  n \equiv Y = \left(\begin{array}{c}\hat{x}_4+i\hat{x}_3\\
  -\hat{x}_2+i\hat{x}_1\end{array}\right) \; , 
\end{equation}
the winding number $k$ of which coincides with the instanton number,
$k = \nu = 1$.

\section{Conclusions}

We have investigated the Laplacian Abelian gauge on the sphere $S^4$
in the background of a single 't~Hooft instanton. This amounts to
solving the eigenvalue problem for the covariant Laplacian in the
adjoint representation.  For any sphere radius $R$ we have determined
the angular dependence and isospin structure of the ground state wave
functions (Higgs fields). Diagonalisation of the latter shows that the
instanton in the singular gauge is in the LAG if $R$ is larger than
the instanton size $\rho$; for the regular gauge the same is true for
$R<\rho$. The gauge fixing procedure thus selects one of the two
instanton configurations, although, in a bundle picture, they
represent the same connection.

It is interesting to note that the situation for the MAG on the sphere
is similar: Singular and regular gauge instantons both satisfy the
differential MAG condition, but the MAG functional $F_\mathrm{MAG}$
picks out one of them in the very same way as $F_\mathrm{LAG}$: for $R
> \rho$ ($R < \rho$) the singular (regular) gauge instanton minimizes
$F_\mathrm{MAG}$ (see Appendix~C). It is, however, a highly nontrivial
task to check whether a given configuration, say the `t~Hooft
instanton, really corresponds to the absolute minimum along its orbit.
In general, one can never be sure that there is no other gauge
equivalent configuration that lowers the functional even further.

The LAG, on the other hand, has the big advantage that the ground
state (and thus the absolute minimum of $F_\mathrm{LAG}$) can be found
explicitly. We have done so for $R=\rho$ and have given qualitative
arguments concerning the angular and radial dependence for the case
$R\neq \rho$.  Apart from the degeneracies and the zeros (which we
have under control), there are no further ambiguities. 

We have found a whole $S^3$ of gauge equivalent configurations
(obtained by \textit{global} SU(2) rotations of $A^{\rm sg}$ similar
to what has been observed in \cite{baulieu:96}) located on the gauge
fixing hypersurface. These are Gribov copies of each other, generated
by both finite and infinitesimal gauge transformations. The latter
give rise to three flat directions in the configuration space along
which the gauge fixing functional does not change. Only one of these
directions is covered by the residual U(1) freedom. The other two are
related to zero modes of the (coset part of the) Faddeev-Popov
operator. We do see no reason why these Gribov ambiguities should not
be present on the lattice. In contrast to the MAG (and related
gauges), however, where gauge fixing is a ``numerical problem of
non-polynomial complexity'' \cite{vink:92}, there are no additional
lattice Gribov copies beyond the denumerable ones we have encountered
in the continuum. This clearly makes the LAG a superior gauge.

Once a ground states is chosen for diagonalisation, additional
obstructions occur in terms of gauge fixing defects caused by the
nodes of the possible ground states. These are the well-known source
for magnetic monopoles in Abelian gauges. We have shown that these
defects \textit{must} be present whenever the LAG background is in a
non-trivial instanton sector. Monopoles, however, only arise for a
particular sphere radius $R=\rho$ and for a particular choice of
ground states. \textit{The generic defects are localised in
space-time} (with codimension 4).  Their topological invariant is the
Hopf index $S^3\rightarrow S^2$. Contrary to monopoles they have
finite action even in the infinite volume limit. One may speculate
that \textit{these} defects condense in the low temperature phase of
QCD, possibly giving rise to a new confinement mechanism.  In view of
the results presented in \cite{tsurumaru:00}, they may as well be
related to the solitonic excitations observed in recent effective
theories for confinement
\cite{cho:99,faddeev:97,faddeev:99,shabanov:99}.

As we have calculated the LAG Higgs field only for a highly symmetric
background, the question arises which features are generic also for
other backgrounds. The degeneracy of the ground state is mainly due to
the matrix structure of the `Hamiltonian'. For a single instanton
background, this was induced by nonvanishing angular momentum (like in
quantum mechanical problems with spin). This should be contrasted with
the case of a trivial background. For the vacuum, $A = 0$, the ground
state obviously has a threefold degeneracy given by the canonical
dreibein $\hat e^a$ in isospace.  The associated constant wave
functions do not have any zeros. We therefore conjecture that Singer's
obstruction \cite{singer:78} against complete gauge fixing is
reflected in the nodes rather than in the degeneracy of the ground
state. To completely settle this issue, a full topological
classification of Higgs field zeros would clearly be helpful.

A natural next step will be to analyse higher instanton sectors and
instanton-anti-instanton pairs. The existence of fermionic zero modes
in the background of Hopf defects is currently being investigated (for
related work see \cite{adam:00} and references therein). Such zero
modes may in the end lead to a relation between confinement and chiral
symmetry breaking. Finally, the dynamical role of Hopf defects in QCD
has to be analysed.

\section*{Acknowledgements}

The authors thank S.~Shabanov and T.~Strobl for enlightening
discussions, P.~de Forcrand for making his lattice results available
prior to publication, and D.~Hansen for a careful reading of the
manuscript. T.V.~gratefully acknowledges the hospitality at the TPI,
University of Jena, where this work was performed. T.H.~thanks
C.~Alexandrou and G.~Burgio for discussions on Laplacian gauges and
acknowledges support under DFG grant WI-777/5-1.

\appendix

\section{Spherical harmonics}
\label{spherharm}

In the following we list the eigenfunctions of $\vec{J}^{\,2}$ and
$\vec{L}^2$ for the three cases of interest (suppressing the two magnetic
quantum numbers labelling the vectors in each multiplet).

\smallskip

\noindent
(i) For $(j,l)=(1,0)$ the spherical harmonics are given by the
canonical dreibein $\hat e^a$ of constant unit vectors,
\begin{eqnarray}
\label{harmonics10}
  Y^{\rm sg}_{(1,0)}=Y^{\rm reg}_{(1,0)}=\left\{
  \left(\begin{array}{c}1\\0\\0\end{array}\right),\:
  \left(\begin{array}{c}0\\1\\0\end{array}\right),\:
  \left(\begin{array}{c}0\\0\\1\end{array}\right)
  \right\}.
\end{eqnarray}
(ii) For $(j,l)=(1/2,1/2)$ there are four eigenfunctions,
all linear in $\hat{x}_\mu$, 
\begin{eqnarray}
  Y^{\rm sg}_{(1/2,1/2)}=\left\{
  \left(\begin{array}{c}\hat{x}_4\\\hat{x}_3\\\hat{x}_2\end{array}\right),\:
  \left(\begin{array}{c}-\hat{x}_3\\\hat{x}_4\\-\hat{x}_1\end{array}\right),\:
  \left(\begin{array}{c}-\hat{x}_2\\\hat{x}_1\\-\hat{x}_4\end{array}\right),\:
  \left(\begin{array}{c}\hat{x}_1\\\hat{x}_2\\\hat{x}_3\end{array}\right)
  \right\}, \label{harmonics12_sg} \\
  Y^{\rm reg}_{(1/2,1/2)}=\left\{
  \left(\begin{array}{c}-\hat{x}_4\\\hat{x}_3\\\hat{x}_2\end{array}\right),\:
  \left(\begin{array}{c}-\hat{x}_3\\-\hat{x}_4\\-\hat{x}_1\end{array}\right),\:
  \left(\begin{array}{c}-\hat{x}_2\\\hat{x}_1\\\hat{x}_4\end{array}\right),\:
  \left(\begin{array}{c}\hat{x}_1\\\hat{x}_2\\\hat{x}_3\end{array}\right)
  \right\}. \label{harmonics12_rg}
\end{eqnarray}
The following remarks are in order.  Obviously, $Y^{\rm reg}$ is
obtained from $Y^{\rm sg}$ upon exchanging $\hat{x}_4 \rightarrow
-\hat{x}_4$.  This is achieved via conjugation with $h$, $Y^{\rm
sg}_{(j=1/2,\,m=1/2)} \sim h^\dagger\, Y^{\rm reg}_{(j=1/2,\,n=1/2)}
\, h$. Note that the `intertwining' gauge transformation $h$ is only
defined up to rotations around the direction of the Higgs field $\phi$
in isospace. It is convenient to combine the members of each (1/2,
1/2) quadruplet into a `four-vector' $Y_\mu$.  Introducing the basis
matrices $\sigma_\mu \equiv (i \sigma^a , \Eins)$, one finds the
relation $Y_\mu^{\mathrm{reg}} = \sigma_\mu Y_\mu^{\mathrm{sg}}
\, \sigma_\mu^\dagger$ for any $\mu = 1, \ldots, 4$. Any component 
$Y_\mu (\hat x)$ vanishes, if $\hat x_\mu = \pm \hat e_\mu$, the $\hat
e_\mu$ denoting the canonical basis of $\mathbb{R}^4$. This means that
the zeros of the quadruplet eigenfunctions are given by two points
located on a three-sphere with fixed radius $r$ (see
Fig.~\ref{loops}).

\smallskip
\noindent
(iii) For the case $(j,l)=(0,1)$ one has three basic eigenfunctions,
now bilinear in $\hat{x}_\mu$,
\begin{eqnarray}
  Y^{\rm sg}_{(0,1)}=\left\{
  \!\!\left(\begin{array}{c}
  \hat{x}_1^2-\hat{x}_2^2-\hat{x}_3^2+\hat{x}_4^2\\
  2(\hat{x}_1\hat{x}_2+\hat{x}_3\hat{x_4})\\
  2(\hat{x}_1\hat{x}_3-\hat{x}_2\hat{x_4})
  \end{array}\right)\!,\!
  \left(\!\!\begin{array}{c}
  2(\hat{x}_1\hat{x}_2-\hat{x}_3\hat{x_4})\\
  -\hat{x}_1^2+\hat{x}_2^2-\hat{x}_3^2+\hat{x}_4^2\\
  2(\hat{x}_2\hat{x}_3+\hat{x}_1\hat{x_4})
  \end{array}\!\!\right)\!,\!
  \left(\begin{array}{c}
  2(\hat{x}_1\hat{x}_3+\hat{x}_2\hat{x_4})\\
  2(\hat{x}_2\hat{x}_3-\hat{x}_1\hat{x_4})\\
  -\hat{x}_1^2-\hat{x}_2^2+\hat{x}_3^2+\hat{x}_4^2
  \end{array}\right)
  \!\!\right\} \\ \label{harmonics01_sg}
  Y^{\rm reg}_{(0,1)}=\left\{
  \!\!\left(\begin{array}{c}
  \hat{x}_1^2-\hat{x}_2^2-\hat{x}_3^2+\hat{x}_4^2\\
  2(\hat{x}_1\hat{x}_2-\hat{x}_3\hat{x_4})\\
  2(\hat{x}_1\hat{x}_3+\hat{x}_2\hat{x_4})
  \end{array}\right)\!,\!
  \left(\!\!\begin{array}{c}
  2(\hat{x}_1\hat{x}_2+\hat{x}_3\hat{x_4})\\
  -\hat{x}_1^2+\hat{x}_2^2-\hat{x}_3^2+\hat{x}_4^2\\
  2(\hat{x}_2\hat{x}_3-\hat{x}_1\hat{x_4})
  \end{array}\!\!\right)\!,\!
  \left(\begin{array}{c}
  2(\hat{x}_1\hat{x}_3-\hat{x}_2\hat{x_4})\\
  2(\hat{x}_2\hat{x}_3+\hat{x}_1\hat{x_4})\\
  -\hat{x}_1^2-\hat{x}_2^2+\hat{x}_3^2+\hat{x}_4^2
  \end{array}\right)
  \!\!\right\}  \label{harmonics01_rg}
\end{eqnarray}
Again, the two sets of eigenfunctions are related via
$\hat{x}_4\rightarrow -\hat{x}_4$ and can most easily be obtained from
case (i) by conjugation with $h$,
\begin{eqnarray}
Y^{\rm sg}_{(j=0,\,m=1)}=h^\dagger\,Y^{\rm reg}_{(j=1,\,n=0)}\,h,\qquad
Y^{\rm reg}_{(j=0,\,n=1)}=h\,Y^{\rm sg}_{(j=1,\,m=0)}\,h^\dagger,
\end{eqnarray}
which, in particular, implies that they \textit{never vanish}.

\section{Feynman-Hellmann theorem and perturbation theory}
\label{FH}

In order to obtain information when $R\neq \rho$, we keep $R$
fixed and vary $\rho$.
We restrict ourselves to the singular
gauge. The $\rho$-dependent part of (\ref{eigeneqn1})
contains two terms,
\begin{eqnarray}
  V_{\rho\,(j,m)}^{\rm sg}(r) \equiv 4e^{-\alpha_R(r)}\left[
  \frac{\rho^2(\vec{J}^2-\vec{M}^2)}{r^2(r^2+\rho^2)}-
  \frac{\vec{T}^2\rho^2}{(r^2+\rho^2)^2}\right].
\end{eqnarray}
The $\rho^2$-dependence of the ground state energy is determined by
the Feynman-Hellmann theorem,
\begin{eqnarray}
  \label{feyn}
  \frac{\pa}{\pa\rho^2}E=
  \frac{\pa}{\pa\rho^2}\langle\phi|H|\phi\rangle=
  \langle\phi|\frac{\pa H}{\pa\rho^2}|\phi\rangle\equiv
  \langle\phi|\frac{\pa V_\rho}{\pa\rho^2}|\phi\rangle.
\end{eqnarray}
For the three angular momentum sectors of interest ($t=1$) we have,
\begin{eqnarray}
  \frac{\pa V_{\rho\,(0,1)}^{\rm sg}(r)}{\pa\rho^2}&=&
  \frac{(r^2+R^2)^2}{R^4}\frac{-4r^2}{(r^2+\rho^2)^3} < 0 \; , \nonumber\\
  \frac{\pa V_{\rho\,(1/2,1/2)}^{\rm sg}(r)}{\pa\rho^2}&=&
  \frac{(r^2+R^2)^2}{R^4}\frac{2(\rho^2-r^2)}{(r^2+\rho^2)^3} \; , \\
  \frac{\pa V_{\rho\,(1,0)}^{\rm sg}(r)}{\pa\rho^2}&=&
  \frac{(r^2+R^2)^2}{R^4}\frac{4\rho^2}{(r^2+\rho^2)^3} > 0 . \nonumber
\end{eqnarray}
According to (\ref{feyn}), these functions have to be integrated with
the positive factor $|\phi|^2\sqrt{g}$. Therefore, the ground state
energies in the first and the third sector are monotonic in $\rho^2$, 
their slopes satisfying
\begin{eqnarray}
  \frac{\pa}{\pa\rho^2}E^{\rm sg}_{(0,1)} < 0 \; ,\qquad
  \frac{\pa}{\pa\rho^2}E^{\rm sg}_{(1,0)} > 0 \; .
\end{eqnarray}
As the energies meet at $R=\rho$ (`level crossing') we conclude,
\begin{eqnarray} 
  E^{\rm sg}_{(0,1)}<E^{\rm sg}_{(1,0)} \quad \mbox{ for } R <\rho \;
  , \qquad 
  E^{\rm sg}_{(0,1)}>E^{\rm sg}_{(1,0)} \quad \mbox{ for } R
  > \rho \; .
\end{eqnarray}
This explains the behaviour of the full lines in Fig.~\ref{er}.

For the sector $(1/2,1/2)$ there is no such simple argument.  Still,
we can compute the slope of $E(\rho^2)$ at the point $\rho=R$ by
simply inserting the known function $\phi$. This amounts to ordinary
perturbation theory in $\delta\equiv \rho^2-R^2$,
\begin{eqnarray}
  H(\rho^2) = H(\delta=0)+\delta\left.\frac{\pa
  H}{\pa\rho^2}\right|_{\delta=0}+O(\delta^2)
  = H_0 + H_{\rm pert} \; . 
\end{eqnarray}
In this way we find a vanishing slope for the sector $(1/2,1/2)$,
\begin{eqnarray}
  \left.\frac{\pa}{\pa\rho^2}E^{\rm sg}_{(1/2,1/2)}\right|_{\rho^2=R^2}
  \sim
  \int_0^\infty \frac{(1-r^2)}{(r^2+1)^7}\,r^5\d r=0 \; . 
\end{eqnarray}
The lowest-lying state of this sector is thus pinched between the
other two, at least for $R\approx \rho$ (cf.\ Fig.\ \ref{er}).

\section{The MAG on the sphere}

In \cite{brower:97b} it has been shown that, due to their particular
Lorentz and isospin structure, both $A^{\rm sg}$ and $A^{\rm reg}$ are
in the MAG when defined on $\mathbb{R}^4$. This still holds true on
$S^4$, where the gauge fixing functional has the values,
\begin{eqnarray}
  F_{\rm MAG}[A]&=&\int
  \sum_{\bar{a}=1}^2 A_\mu^{\bar{a}} A_\nu^{\bar{a}} \, g^{\mu\nu}\sqrt{g}\,
  \, \d^4 x\nonumber\\
  &=&
  \frac{16\pi^2R^4 [R^4-2R^2\rho^2\ln(R^2/\rho^2)-\rho^4]}
  {\rho^2(R^2-\rho^2)^3}\times
  \left\{\begin{array}{ll}
  1&\mbox{for }A^{\rm sg}\\
  R^2/\rho^2&\mbox{for }A^{\rm reg}
  \end{array}\right. \; . 
\end{eqnarray}
Obviously, $F_{\rm MAG} [A^{\rm reg}] = (R^2/\rho^2) \, F_{\rm MAG}
[A^{\rm sg}]$, so that for $R > \rho$ ($R < \rho$) the singular
(regular) gauge is singled out.

\bibliographystyle{../../h-physrev}
\bibliography{../../gauge}

\end{document}